# Computational Optimization of MnBi to Enhance Energy Product

Tula R Paudel[1a)], Bhubnesh Lama[1], Parashu Kharel[2]
[1] Department of Physics, South Dakota School of Mines and Technology, Rapid City, SD 57701
[2]Department of Physics, South Dakota State University, Brookings, SD 57707

ABSTRACT

High energy density magnets are preferred over induction magnets for many applications, including electric motors used in flying rovers, electric vehicles, and wind turbines. However, several issues related to cost and supply with state-of-the-art rare-earth-based magnet necessities development of high-flux magnets containing low cost, earth-abundant materials. Here, we demonstrate the possibility of tuning magnetization and magnetocrystalline anisotropy of one of the candidate materials, MnBi, by alloying it with foreign elements. By using the density functional theory in the high-throughput fashion, we consider the possibility of alloying MnBi with all possible metal and non-metal elements in the periodic table and found that MnBi-based alloys with Pd, Pt, Rh, Li, and O are stable against decomposition to constituent elements and have larger magnetization, energy product compared and magnetic anisotropy compared to MnBi We consider the possibility of these elements occupying half and all of the available empty sites. Combined with other favorable properties of MnBi, such as high Curie temperature and earth abundancy of constituents elements, we envision the possibility of MnBi-based high-energy-density magnets.

[a)] Author to whom correspondence should be addressed:tula.paudel@sdsmt.edu

MnBi, a member of Mn-A alloys, where A can be any element with $3^{+}$ oxidation state as Al, Ga, In, has potential to be a good magnet [1–4]. The compound posses considerable magnetocrystalline anisotropy energy of ( ~0.163 meV/uc) [5] that increases with temperature up to 553 K [6] and a high coercive field 17kOe [7] at room temperature; such a combination leads to high energy product (BH)max of 17 MGOe [5] interesting for magnetic applications. Additionally, the compound has a large Curie temperature of 628K [1] and well suited for high-temperature applications. Even with such favorable properties, the MnBi and MnBi-based compounds suffer from low saturation magnetization [8]. Various ideas, including defect engineering [2,9–14], exchange coupling with the soft magnets [8,11,15–21], and microstructural engineering [22–26], have been tested to improve the energy products in these compounds. However, they have so far yielded mixed results. While most of the doping with various foreign elements and microstructure refinement have shown an increase of coercivity at the expense of magnetization [2,3,9,27], Sn doping [3] is found to increase magnetization and magnetic anisotropy, leading to an increase in energy products.

In this manuscript, we test an alternative approach of alloying MnBi with foreign elements to enhance magnetization and magnetocrystalline anisotropy leading to enhanced energy product $(BH)_{max}$, of the materials. We use the high throughput density functional theory for screening suitable elements to form an alloy with MnBi. The structure of MnBi offers clues on why alloying may be possible in MnBi. The low-temperature phase of the MnBi crystallizes in hexagonal NiAs type structure (Space Group No. 194), as shown in Fig. 1. The lattice vectors of such a crystal are $A_1 = 1/2\, a\hat{x} + \sqrt{3}/2\, a\hat{y}$ ; $A_2 = 1/2\, a\hat{x} - \sqrt{3}/2\, a\hat{y}$ and $A_3 = c\hat{z}$, where $a$ and $c$ are lattice constants. Mn occupies 2a Wyckoff's positions with $D_{3d}$ site symmetry: 0 and $(1/2\, A_3)$; Bi occupies 2c Wyckoff's positions with $D_{3h}$ symmetry $(1/3\, A_1 + 2/3\, A_2 + 1/4\, A_3)$ and $(2/3\, A_1 + 1/3\, A_2 + 3/4\, A_3)$ while the other high symmetry Wyckoff's 2d positions : $(2/3\, A_1 + 1/3\, A_2 + 1/4\, A_3)$ and $(1/3\, A_1 + 2/3\, A_2 + 3/4\, A_3)$ are empty. We incorporate foreign elements to these sites and search for the elements that increase magnetization and coercivity. We consider two cases for each elemental (*el*) insertion. i) only one of two vacant sites is filled with foreign elements (half-filled case); the other half-filled configuration obtained by filling the rest empty position is expected to behave similarly because of the similarity of lattice symmetry and the local environment surrounding the elements when placed in that positions, and ii) both of two vacant sites are filled (full-filled case). Our strategy to find the suitable element for alloying is to first evaluate the formation energy of each alloy to find the stable one; second, determine saturation magnetization of stable alloys; and finally, calculate magnetic anisotropy energy of alloys that are stable and have magnetization larger than undoped MnBi.

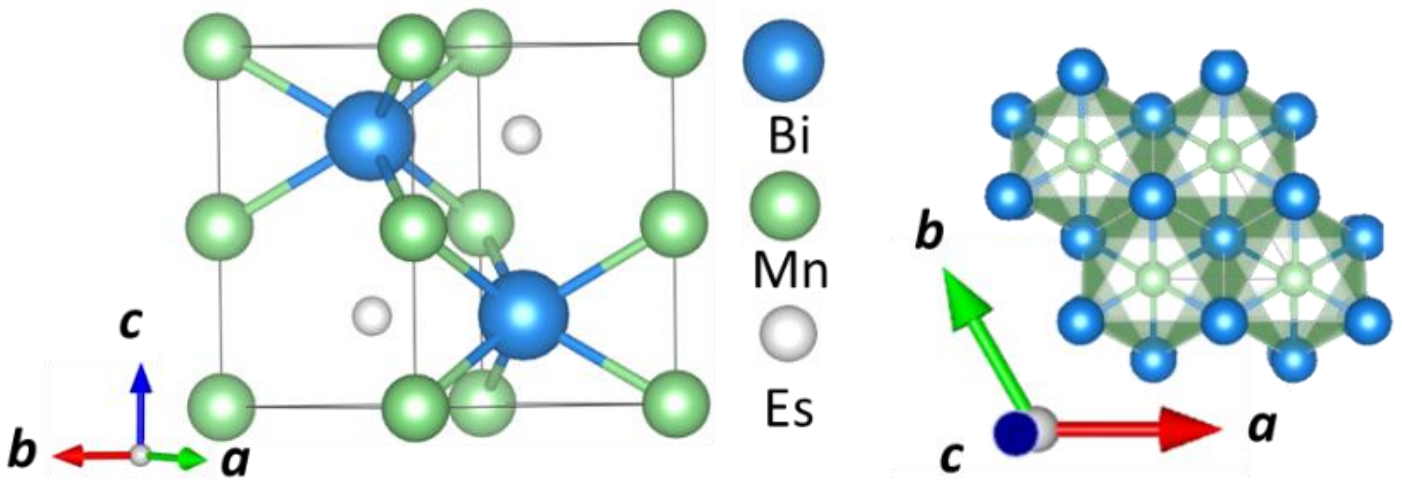

Figure 1: NiAs type low-temperature structure of MnBi side view in the left and top view on the right.

For the stable alloy, the formation energy, $\Delta H$, must be negative, which ascertains energy gain by alloying. The formation energy of the MnBi:el is evaluated as $\Delta H = E(\text{ MnBi: el}) - E(MnBi) - E(el)$. Here $E(MnBi)$ and $E(el)$ are energies of MnBi and ground state energy of element, el. Similarly, E(MnBi:el) is the energy of the ground state configuration of the alloy. We determine the ground state energy of the alloy by comparing the energies of two configurations: first, ferromagnetic (FM) configuration in which magnetic moment of alloying elements aligns parallel to the magnetic moments of Mn; and second antiferromagnetic (AFM) configurations in which the magnetic moment of alloying element is aligned anti-parallel to the magnetic moments of Mn, determine their energies. The calculated formation energy from these calculations is likely to be in the order of ~0.1 eV/formula unit, corresponding to ~10 KJ/mol [28].

| 1 IA | 2 IIA | 3 IIIB | 4 IVB | 5 VB | 6 VIB | 7 VIIB | 8 | 9 VIIIB | 10 | 11 IB | 12 IIB | 13 IIIA | 14 IVA | 15 VA | 16 VIA | 17 VIIA |
|---|---|---|---|---|---|---|---|---|---|---|---|---|---|---|---|---|
| 1.71 **H** 3.32 | | | | | | | | | | | | | | | | **H** |
| 0.64 **Li** -1.27 | 1.42 **Be** 2.77 | | | | | | | | | | | 2.24 **B** 6.54 | 4.75 **C** 9.42 | 3.08 **N** 5.35 | 0.1 **O** -0.72 | -1.33 **F** -2.7 |
| 2.65 **Na** 0.94 | 1.52 **Mg** 0.9 | | | | | | | | | | | 1.29 **Al** 2.37 | 1.9 **Si** 3.12 | 3.47 **P** 2.54 | 1.72 **S** 1.84 | 1.81 **Cl** 2.02 |
| 2.77 **K** 3.23 | 0.72 **Ca** 0.1 | 0.16 **Sc** 0.74 | 0.39 **Ti** 0.73 | 0.9 **V** 1.75 | 1.19 **Cr** 2.27 | 1.15 **Mn** 2.09 | 0.97 **Fe** 2.35 | 0.41 **Co** 1.37 | 1.12 **Ni** 1.03 | 0.75 **Cu** 1.33 | 2.5 **Zn** 1.68 | 1.31 **Ga** 1.74 | 1.73 **Ge** 2.87 | 1.83 **As** 2.31 | 1.86 **Se** 1.95 | 2.55 **Br** 2.99 |
| 3.81 **Rb** 4.98 | 1.5 **Sr** 0.88 | 0.68 **Y** -0.24 | 0.7 **Zr** 0.86 | 1.35 **Nb** 2.89 | 1.76 **Mo** 4.0 | **Tc** | 1.94 **Ru** 1.83 | 0.04 **Rh** -0.02 | -0.06 **Pd** -0.46 | 1.11 **Ag** 1.67 | 1.62 **Cd** 2.42 | 1.93 **In** 2.65 | 2.12 **Sn** 2.78 | 2.67 **Sb** 2.6 | 4.64 **Te** 2.79 | 3.64 **I** 4.58 |
| | 2.49 **Ba** 2.72 | 1.11 **La** 0.12 | 0.97 **Hf** 1.42 | 1.79 **Ta** 4.03 | 2.59 **W** 5.65 | 2.31 **Re** 4.91 | 1.85 **Os** 3.76 | 1.88 **Ir** 1.28 | 0.09 **Pt** 0.06 | 0.95 **Au** 0.94 | **Hg** | **Tl** | 2.58 **Pb** 3.24 | 4.47 **Bi** 3.2 | **Po** | **At** |

Figure 2. The formation energy of the MnBi alloy with the various elements. The two numbers above and below each element in the box represent the formation energy, $\Delta H$, of half-filled (MnBi:$el_{1/2}$) and full-filled (MnBi:el) alloy with the element (el). The isolated box shows the legend. The highlighted elements form a stable alloy with MnBi.

Figure 2 shows the formation energy of the alloy with the various elements in the periodic table. Out of all those calculations, the attractive alloys are the ones with the negative formation energy. In the half-filled case, only F and Pd doped alloys have negative formation energies. These are marked with light color in Figure 1. Similarly, in the full-filled case, F, Pd, Li, O, Y, Rh doped alloys also have negative formation energies, indicating the alloys to be stable with respect to decomposition to the elemental phase. Other half-filled alloys with Rh, Pt, and Sc have formation energies less than 0.2 eV; these dopants only partially occupy these vacant sites. We calculate site occupancy $N_{occ}/N = \exp(-\Delta H/K_B T)$ at 500 K, where $N$ is the number of available sites, $K_B$ is the Boltzmann constant, and T is the temperature. We found that Rh, Pt, and Sc occupy 40%, 25%, and 2% of empty sites, respectively. Similarly, full-filled MnBi alloys with Pt, Ca, and La has slightly positive formation energy (<0.2 eV) and only partially occupy the empty sites: Pt occupies 23%, Ca occupies 9%, and La occupies 6% of the empty sites. When $\Delta H$ > 0.2 eV interstitial sites of MnBi remains empty.

Table I: Change in saturation magnetization, anisotropy, volume formation, $c/a$, estimated Curie Temperature ($T_c$) and change in energy product, $\Delta(BH)_{max}$(%) as a function of alloying elements (el) in half-filled MnBi: $el_{1/2}$ and full-filled MnBi:el alloys.

| Case | **ΔM(%)** | ΔAE($MJ/m^3$) | $\Delta V_F$($Å^3$) | *c/a* | $T_c$ (K) | $\Delta(BH)_{max}$(%) |
|---|---|---|---|---|---|---|
| MnBi | 0 | 0 | | 1.33 | 580 | 0 |
| MnBi:Li | 9 | 4 | -1.27 | 1.24 | 190 | 18 |
| MnBI:O | 9 | 5 | -0.72 | 1.05 | 555 | 18 |
| MnBi:$F_{1/2}$ | -2 | - | | | | |
| MnBi:F | -7 | - | | | | |
| MnBi:$Sc_{1/2}$ | -9 | - | | | | |
| MnBi:Y | -7 | - | | | | |
| MnBi:$Rh_{1/2}$ | 8 | 3.0 | -1.53 | 1.33 | 522 | 16 |
| MnBi:Rh | 14 | 0.92 | -0.02 | 1.21 | 312 | 32 |
| MnBi:$Pd_{1/2}$ | 9 | 20.7 | -0.46 | 1.33 | 457 | 18 |
| MnBi:Pd | 14 | 0.33 | -0.46 | 1.22 | 327 | 30 |
| MnBi:Pt | 10 | 4.79 | 0.06 | 1.20 | 417 | 21 |

All the alloys with negative formation energy are found to have a larger volume compared to the MnBi. The lattice volume increases, but modestly, according to the increase in the atomic radius [29]. The smallest volume increase is observed in F (1%), and the largest volume increase is observed in the case of Y (42%). Volume increases more in the full-filled than half-filled case; however, the increase is not linear; for example, the volume increase of full-filled MnBiRh is ~ 19%, whereas that of half-filled MnBi$Rh_{1/2}$ is ~ 14%. This nonlinear increase in volume indicates hybridization between the alloying elements and MnBi while both empty sites are occupied. Despite the increase in volume, the volume formation, which is defined in a similar way as the formation energy, $\Delta V_F = V(MnBi{:}nel) - (V_{MnBi} - nV_{el})$, where $n$ is number of volume $V_{el}$ and $V_{MnBi}$ is the volume of MnBi, is generally negative (except for MnBi: Pt) as shown in Table I. The negative $\Delta V_F$ also indicates hybridization between the valence orbitals of alloying elements with valence orbitals of Mn and Bi in MnBi.

Having found the stable alloys, we next analyze their magnetic properties. Table I. shows changes in magnetization with respect to the 3.54 $\mu_B$/Mn of bulk MnBi. Out of the alloys with negative formation energy alloying with F, Sc, Y reduces the net moment, where alloying with Li, O, Rh, Pd, and Pt increases net moment by up to 14% compared to the undoped case. In the case of F doping, F-*p* orbital hybridizes strongly with Bi-*p* orbitals (Fig. 2c), which couples to Mn-d orbitals antiferromagnetically like that in bulk MnBi, thereby reducing the net moment. In the case of Sc and Y, their 3*d*-states, as shown in Fig 2d and Fig 2e, mainly lie in the conduction band; thus, they hybridize with Mn minority d-states, inducing small moments in themselves and reducing moments in Mn, and the moments in Y and Sc couple antiferromagnetically with Mn moments further reducing the overall magnetizations.

Alloying with Rh, Pd, Pt increases the net magnetization of the system. 4*d* orbitals of the Rh (Fig. 2f) and Pd (Fig 2g) and 5d orbitals of the Pt (Fig 2h) lies mainly in the valence band and hybridize strongly with the majority spin-channel of Mn, which induces small magnetic moments on themselves and enhances magnetic moments on Mn. These moments on Rh, Pd, and Pt, however, couples ferromagnetically with Mn moments. This is opposite to interaction between Mn at the regular sites and interstitial sites, which couples antiferromagnetically [30,31] or interaction between Sc, Y in interstitial sites and Mn at regular sites, as we discussed previously. The difference comes from the large exchange splitting of Mn-d bands leading to almost no interaction between electrons with different spins located at the different sites. However, in Rh, Pd, and Pt, even though intra-atomic exchange interaction leads to *d*-orbitals splitting according to spin like that in Mn, the splitting is much smaller. As a result, bands corresponding to both spin channels are occupied though there is a slight shift between them, which results in a small magnetic moment. Additionally, they have a significant overlap with Mn *d*-orbitals occupied by the majority spin channel, resulting in ferromagnetic direct exchange.

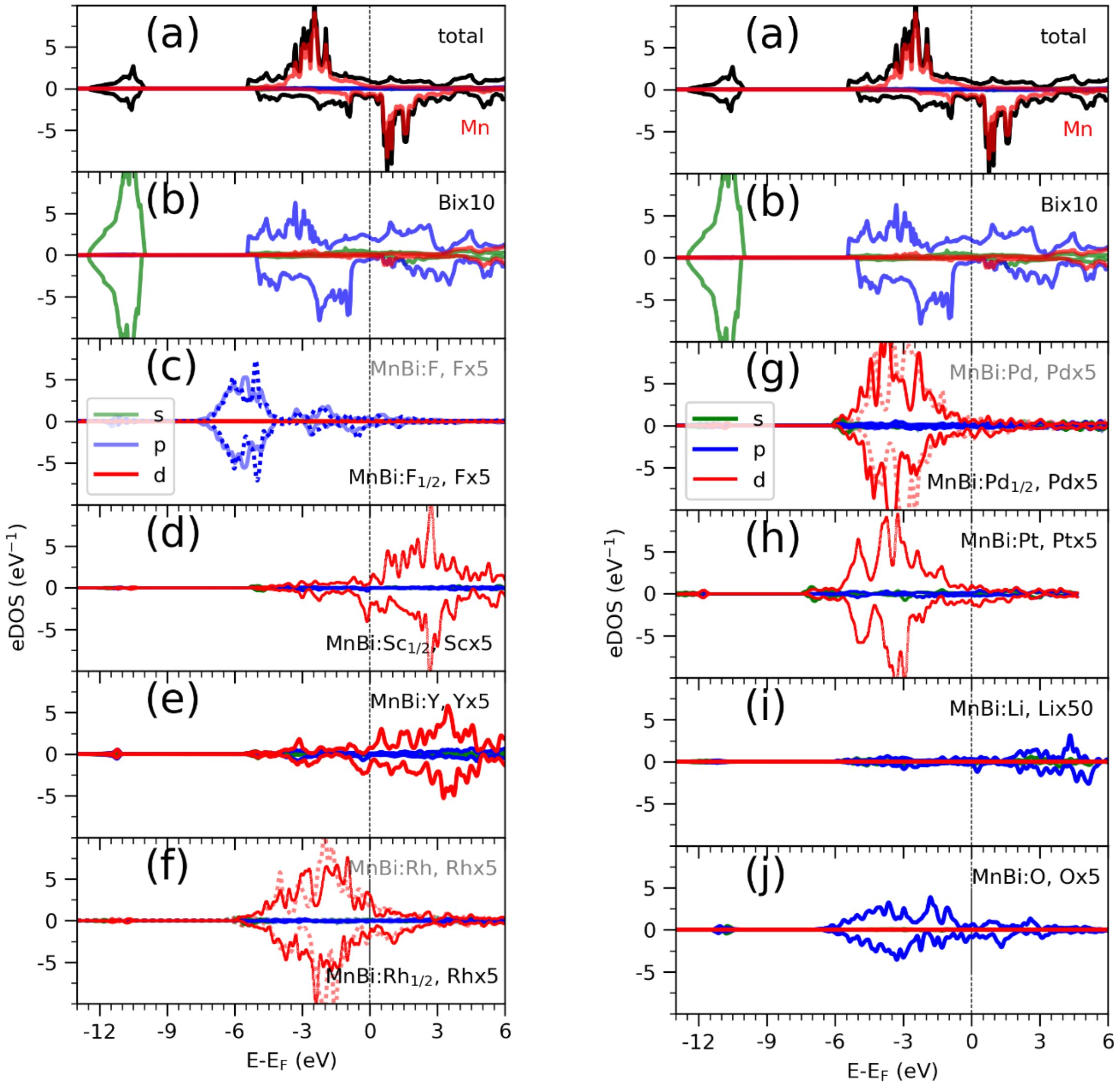

Figure 2: Atom, spin and orbital resolved density of states (DOS) of MnBi (a) and (b) and its alloying elements (el) in stable full-filled MnBi:el and half-filled MnBi:el$_{1/2}$ as indicated in left of panels (c) to (j). In the cases (F, Rh and Pd), where both half-filled and full-filled alloys are stable, the DOS of element in full-filled alloy is shown with dotted line in the background of solid line that corresponds to DOS of the element in half-filled case. Colors red, blue, and green corresponds to $d$, $p$ and $s$ orbitals respectively. $\times x$ in c-f and g-j corresponds to the enlargement factor for DOS for better visualization.

Alloying group II elements like Li and O also increases overall magnetization. The *s*-states of Li and $p-$states of O lies on the opposite side of the Fermi energy structure: the O-*p* states lie primarily in the valence band as shown in Fig. 2j, while Li-*s* states lie in the conduction band as shown in Fig. 2i. As a result, the hybridization between Li-s states and Mn-*d* states, and O-*p* and Mn-*d* states is spin-dependent; the oxygen states hybridize mostly with the Mn-d majority states, while Li-s states hybridize mostly with Mn-d minority states leading to the enhancement of Mn moments and development of small moments in Li and O themselves. The magnetic moments of Li and O also couple ferromagnetically because of direct exchange with Mn moments and help increasing overall magnetization. We verify this hypothesis by explicitly comparing the energy of the system with the moments of Mn and moments of Li and O coupling ferromagnetically and antiferromagnetically and found that the ferromagnetically coupled system has lower energy. Using the energy difference $\Delta E = E_{FM} - E_{AFM}$ we estimate the Curie temperature in the mean-field approximations by using $T_c = \Delta E/3$ [31]. For the bulk MnBi, we calculate $\Delta E$ =150 meV/Mn and $T_c \approx$ 580 K, which is resonably close to extrapolated experimental value of 775K [6]; The structural transition at 628K hinder measuring real $T_c$ of MnBi. Using the same approach, we eavaluate the $T_c$ of stable MnBi alloys with enhanced magnetizations. The second to last column of the table I shows the calculated results, which shows that $T_c$ reduces in general upon doping. This is consistent with the fact that upon doping the volume increases, c/a decreases and hence the first exchange parameter $J_1$ , which contribute the most to total exchange $J,$ decreases due to reduced overlap between the orbital along out-of-plane directions. This estimate ignore the contribution to the Tc second and higher other exchange and should represent the lower bound.

Next, we calculate change (*Δ*) in magnetic anisotropy energy MAE, *Δ*MAE = MAE (MnBi:el/el$_{1/2}$) - *MAE(MnBi),* in stable alloys with magnetization larger than that of bulk MnBi. MAE is computed by taking the energy difference between the system with magnetization pointing along [100] direction and the system with magnetization pointing along [001] direction, $MAE = E(M||[100]) - E(M||[001])$. The results in Table I show that anisotropy energy increases upon alloying compared to MnBi (~1MJ/m$^3$) in all cases. Since the anisotropy energy $E = K_1 \sin^2\theta + K_2 \sin^4\theta$ , where $\theta$ is the angle between easy axis [001] and the direction of magnetization, and $K_2 << K_1$ ΔMAE roughly corresponds to change in $K_1$. We note that magnetization of MnBi alloy with Rd and Pd$_{1/2}$ points along [001] direction, similar to the bulk MnBi, the magnetization of MnBi alloy with Li, O, Pd, and Pt changes to in-plane [100] direction. Qualitatively, such changes in MAE and easy axis are related to subtle changes in band distribution near the Fermi level and may require separate detailed investigations. The magnetic anisotropy is a relativistic effect driven by spin-orbit coupling. As the spin-orbit coupling energy is much smaller than the bandwidth of Mn-3*d* states, perturbative expansion of energy can be used to explain the spin-orbit coupling effect. The lowest non-zero correction to the energy in such expansion includes the term proportional to $1/(E_{unocc} - E_{occ})$ where $E_{occ\ (unocc)}$ is the energy of occupied (unoccupied) state around Fermi energy. The non-zero DOS of alloying elements near the Fermi level indicates the possibility of such changes near Fermi levels.

Finally, we estimate the theoretical value of maximum energy product $(BH)_{max}$ assuming applied field $H$ much smaller than magnetization, M. In this approximation, magnetic induction field $B = 4\pi M$ in CGS units, and $M_r = M_s$ when magnetic hysteresis loop is almost rectangular. The $(BH)_{max} = B_r^2/4$, where the remnant field $B_r = 4\pi M_r = 4\pi M_s$ [32]. The formula allows us to estimate the energy-product using only the intrinsic properties such as actual magnetization (emu/cc) and homogeneous sample without using coercivity that relies on the extrinsic factors such as shape and size of the sample. The last column of Table I shows the calculated percentage change in energy product to calculated energy product of 20MGOe, which is similar to the value of 17 MGOe [33,34] measured experimentally. The energy product enhances when MnBi is alloyed with Pd, Pt, Rh, Li, and O per the change in magnetizations.

In summary, we used the first-principles density functional theory to find an alloy with the increased magnetization and anisotropy. We predict MnBi alloy with Pd, Pt, Rh, Li, and O are stable against decomposition to constituent elements and have larger magnetization and anisotropy compared to MnBi and have a high energy product. Magnetic easy axis of MnBi alloy with $Pd_{1/2}$ and Rh remains the same as bulk MnBi and lies out-of-plane direction, while the magnetic-easy-axis of MnBi alloy with Li, O, Pd and Pt lie in-plane. Pd can rotate the magnetic-easy-axis [35] from in-plane to out-of-plane depending upon the percentage of Pd that would be incorporated in MnBi. We anticipate this comprehensive study of MnBi alloy spurs more theoretical and experimental study. We note that the alloying of these elements with MnBi may require using out-of-equilibrium growth methods as their formation energies are relatively small.

## Computational Methods

We employ the projected augmented wave (PAW) [36] method for the electron-ion potential and the gradient density approximation (GGA) for exchange-correlation potential, as implemented in the Vienna *ab-initio* simulation package (VASP) [37,38] with the recommended set of pseudopotentials. The calculations are carried out using the kinetic energy cutoff of 340 eV and 6×6×1 *k*-point mesh for Brillouin zone integration of pseudocubic unit cells. The k-points are scaled according to size. We fully relax ionic coordinates with the force convergence limit of 0.001 eV/atom. For the anisotropy calculations, we include 0.1an additional onsite Coulomb interaction Hubbard (U-J) parameter of 3eV to Mn-3d states so that the sign of magnetocrystalline of bulk MnBi is consistent with the experiments and includes spin-orbit coupling terms explicitly.

## Acknowledgments

This work was supported by South Dakota-NASA EPSCoR Research Initiation Grant no 80NSSC19M0063. Computations were performed utilizing the Holland Computing Center at the University of Nebraska, Lincoln, and the Gamow Computing Cluster at the Department of Physics, South Dakota School of Mines and Technology. The authors acknowledge discussion with Alex Leary, Research Materials Engineer at the NASA Glenn Research Center, Cleveland OH.

## Availibily of Data

The data that support the findings of this study are available from the corresponding author upon reasonable request.